# Evaporating drops of alkane mixtures

G.Guéna, C.Poulard, A.M.Cazabat*

Collège de France, 11 place Marcelin Berthelot, 75231 Paris Cedex 05

- *Corresponding author

Abstract:

Alkane mixtures are model systems where the influence of surface tension gradients during the spreading and the evaporation of wetting drops can be easily studied. The surface tension gradients are mainly induced by concentration gradients, mass diffusion being a stabilising process. Depending on the relative concentration of the mixture, a rich pattern of behaviours is obtained.

## I. Introduction

Heat transfer controlled by the evaporation of liquids on solid substrates, is the subject of many experiments and numerical analyses [1-4]. On the other hand, most industrial coating processes use solutions of the active compounds in volatile solvents. In both cases, surface tension gradients result from thermal and/or concentration gradients. The surface flow may significantly interplay with the heat transfer or coating process, possibly leading to film rupture and dendritic instabilities [1-9]. The behaviour of the contact line is of specific interest from the theoretical point of view, because both the evaporation flux and the dissipation are singular in a hydrodynamic analysis [10-11].

We are interested in situations where the gradients are merely due to the evaporating process, and not driven externally. In that case, there is a significant difference between surface tension gradients induced by thermal or concentration gradients. Thermally induced surface tension gradients are usually weak, and thermal diffusion, which is a stabilising process, is fast. Moreover, thermal exchanges with the substrate come into play. On the contrary, concentration-induced surface tension gradients are usually large, mass diffusion is slow, and there is no mass exchange with the substrate.

The present paper is devoted to the evaporation of drops composed of binary alkane mixtures. This is a simple case if compared with polar mixtures previously studied by Marmur et al. [12-13], because the interactions in alkanes are purely dispersive, the mixtures are ideal ones, and complete wetting is always achieved. The same systems have already been studied in another geometry with a liquid reservoir [14-15], then mimicking the well-known "tears of wine" problem [16-17].

## II. Materials, methods and a summary of results for pure alkanes

The substrates are oxidised silicon wafers, made hydrophilic by proper cleaning ("piranha" solution: 1/3 $H_2O_2$, 2/3 $H_2SO_4$). The liquids used are alkanes, from nonane to hexane (Aldrich, >99% pure), most of the results being obtained with the heptane/octane mixture. Pure alkanes as well as their mixtures wet the substrates completely.

The relevant physical parameters of the pure alkanes at 20°C are given in table I.

| liquid | viscosity $\eta$ x$10^{-3}$ Pa.s | surface tension $\gamma$ x$10^{-3}$ Nm$^{-1}$ | evaporation parameter $j_0$ x$10^{-10}$ m$^2$s$^{-1}$ |
|---|---|---|---|
| hexane | 0.326 | 18.4 | 34 |
| heptane | 0.409 | 20.5 | 16 |
| octane | 0.542 | 21.8 | 3.2 |
| nonane | 0.711 | 23 | 1 |

When deposited on the substrate, the drops first spread and the radius $R(t)$ of the wetted spot increases. Then spreading and evaporation balance and $R(t)$ goes to a maximum value $R_{\max}$.

Later on, evaporation becomes dominant, the radius decreases, and the drop ultimately disappears at a time $t = t_0$.

The dynamics of drops is followed under microscope at low magnification, in normal atmosphere, with protections against air draft. The drops are always very flat, therefore equal thickness interference fringes are easily observed and allow us to get information on the profiles.

Our previous work has dealt with the evaporation of the pure alkanes on the same kind of substrates [11, 18]. During the retraction of the drops, the radius $R(t)$ scales as $(t_0 - t)^y$, where the exponent $y$ is slightly less than 0.5 (between 0.48 for hexane and 0.44 for nonane).

Such a behaviour is reminiscent of the one of spherical aerosol droplets evaporating slowly in a gas phase, in which case the exponent is exactly 0.5 [19-20]. The reason why is that evaporation is controlled by the diffusion of the vapour molecules in the gas phase. Just above the free interface, the concentration $c$ of the vapour is equal to the value at saturation $c_{sat}$, therefore a constant if thermal and curvature effects are negligible. At infinity, $c$ is also fixed and equal to zero. With these fixed boundary conditions, a quasi-stationary solution is reached rapidly. In such a diffusive, stationary process, the laplacian $\Delta c$ is equal to 0.

Given that the evaporation is proportional to $\nabla c$ taken at the interface, the total rate of evaporation of the drop is proportional to the drop radius (and not to the drop area, as it would be the case in vacuum). This feature is characteristic of diffusion limited evaporation and leads to an exponent 0.5 in the case of spherical drops, in good agreement with experiments.

For sessile drops, the presence of the substrate, and the specific processes occurring at the moving contact line make the problem more complex. Now, both radius and contact angle have to be considered. However, evaporation still proceeds by the stationary diffusion of the vapour molecules in the gas phase, and the laplacian $\Delta c$ is still equal to 0. The total rate of evaporation of the drop is proportional to the radius of the wetted spot, and the exponent $y$ is close to 0.5. These evaporating drops are always close to spherical caps, with small contact angles, of the order of 0.04 rd or less.

The situation is significantly different with mixtures.

**III. Experimental results for alkane mixtures**

A remarkable feature is that, for a given initial volume, a drop of mixture may evaporate faster than a drop of the most volatile pure alkane, see **figure 1a**. As the rate of evaporation of a drop increases with the size of the wetted spot, this corresponds to the fact that the maximum radius $R_{max}$ for the mixture is usually larger than the ones of pure alkanes, see **figure 1b**. This non-monotonous dependence of $R_{max}$ with the composition of the mixture is due to the presence of outwardly directed surface tension gradients, which accelerate the spreading process, see **figure 1c**. One may also record the maximum radius as a function of deposited volume for various initial concentrations of the mixture, see **figure 1d**.

A more detailed observation reveals a very rich pattern of behaviour. We observe at short time after deposition (1-10s) that, for any mixture at any concentration, the shape of the drop is distorted. This corresponds to the change in the spreading rate due to arising surface tension gradients (see figure 1c). At intermediate concentrations, more or less complete breaking is observed during the receding motion between the central part of the drop and an outer ring, whereas no breaking is observed at low and high concentration. In the case where the drop splits in two parts, the centre evaporates faster than the ring at lower concentrations, and conversely, see **figure 2a,b**.

In the present paper, we will focus on the heptane-octane mixture, and give some trends for other ones. Heptane evaporates four times faster than octane, and its surface tension is smaller than the one of octane (see table I). The molecule sizes are approximately the same.

In the ranges of concentration where the drops are "stable", it is possible to define and measure their radius as a function of time. The data are plotted on **figure 3** for a given initial volume $V$ of the drops as a function of $t_0 - t$, together with the data for pure alkanes. For $V$ = 0.4µL, the range of initial concentrations $x_0$ where drops are "unstable" is roughly 0.065 < $x_0$ < 0.17. Here, $x_0$ is defined either as the initial relative volume of the more volatile alkane or as the relative initial number of molecules.

One sees immediately that the curves at "low" concentration, $x_0$ < 0.065 differ qualitatively from the ones at "large" concentration, $x_0$ > 0.17. At low initial concentration, the curves merge with the one of the pure, less volatile alkane within the drop life. This means that the more volatile alkane has completely disappeared, together with any concentration gradient, before the end of the evaporation process. This is no longer the case for large initial concentrations $x_0$ > 0.17, in which case the gradient is obviously present during the whole process.

The same behaviour is observed with the other mixtures, see **figure 4** for octane-hexane and heptane-nonane.

Unstable drops can be characterised qualitatively by the time evolution of the profile. The observed behaviour together with the schematic evolution of shapes is summarised on **figure 5**. Stable drops at "low" concentration are observed in the $\alpha$-domain of the diagram, see **figure 6**. At larger concentration and / or volume, unstable drops are obtained, where the centre of the drop disappears before an external ring. If concentration is increased, the external ring disappears first. Then stable drops are again observed, corresponding to the "large" concentrations in figures 3-4. This is the $\beta$-domain of the diagram, see **figure 7**. Unstable drops correspond to the $\gamma$-domain. Corresponding pictures are shown on figure 2a,b. Very small drops are stable whatever the initial concentration is, which means that $\alpha$- and $\beta$-domains merge at the bottom of the diagram on figure 5.

Even for stable drops, the profiles are far from being always spherical caps. They result from a complex interplay between evaporation, surface tension gradients, hydrodynamic flow, mass diffusion, and capillary forces. It is clear that both capillary forces, which tend to reduce the drop area when the surface tension is constant, and mass diffusion will be dominant for very small drops, and promote stability. This explains why small drops are stable, and also why the boundary between $\alpha$- and $\gamma$-domains depends both on concentration and drop size. Saying more needs a specific analysis.

**IV. Equations and simplified models**

The aim of this part is not to write and solve exact equations, which is largely out of the scope of the paper, but rather to introduce explicitly the specific parameters relevant for the problem, and the typical approximations involved. After a brief reminder of the equations available for pure alkanes, the case of mixtures will be addressed.

### IV.1. General equation for an evaporating drop [11].

The starting point for the evolution of drops is the local conservation equation (radial symmetry is assumed):

$$\frac{\partial h}{\partial t} + \nabla(hU) = -J(r) \qquad (1)$$

Here, $h \equiv h(r,t)$ is the local thickness, $r$ the distance to the drop axis, $U$ the velocity averaged over the thickness, and $J(r)$ the evaporation rate per unit area of the substrate.

$$U(h,t) = \frac{h^2}{3\eta}\nabla(\gamma\Delta h + \Pi(h)) + \frac{h}{2\eta}\nabla\gamma \qquad (2)$$

$\eta$ is the viscosity of the liquid, $\gamma$ the surface tension. There are two main contributions[1] to the pressure: $\gamma\Delta h$ is the capillary term (the Laplace pressure), and $\Pi(h)$ the disjoining pressure, which accounts for the specific properties of very thin films [21]. The last term in equation (2) is the Marangoni flow, and shows up in the presence of a surface tension gradient.

For a **homogeneous** liquid, assuming a diffusion-controlled, quasi-stationary evaporation process, and considering that the slope of the interface, and noticeably the contact angle, are very small, the evaporation rate can be written as [22-25]:

$$J(r) = \frac{j_0}{R\sqrt{1-\left(\frac{r}{R}\right)^2}} \qquad (3)$$

where $R$ is the radius of the drop. As already mentioned, the rate of change of the drop volume, obtained by integrating the conservation equation over the drop's surface, is proportional to $R$, like in aerosols. However, in the case of sessile drops, the rate of evaporation is not uniform, in contrast with aerosols. $J(r)$ is maximum at the edge, because the molecules leaving the liquid interface there have a lesser probability to come back to the drop during their diffusive motion inside the gas phase.

---

[1] The hydrostatic pressure is negligible and has been omitted.

Solving the dynamic equation requires to analyse explicitly the behaviour of various terms (interface curvature, disjoining pressure and evaporation rate) at the moving contact line, because they all diverge in a macroscopic description. A regularisation procedure has been proposed by M. Ben Amar and co-workers, and accounts satisfactorily for the data obtained with alkanes [11]. In the following, we shall therefore ignore the various divergences at the contact line, because the solution of the problem is available elsewhere. Noticeably, the disjoining pressure will no longer be considered.

### IV.2. The case of mixtures.

The case of mixtures is necessarily complex because, after some time, the local concentration in the drop depends explicitly on position and time, due to the variation of evaporation rate over the interface. The places where $J(r)$ is larger are depleted in the most volatile alkane. As the most volatile alkane also has the lowest surface tension, surface tension gradients result, directed towards the places where $J(r)$ is larger.

An explicit resolution of the equations is out of question, therefore we shall discuss only very specific, simple cases and propose empirical arguments for the observed behaviour.

Equations (1-3) must be first rewritten in the case of mixtures.
A first, plausible approximation is to assume a **complete mixing of the two alkanes over the thickness of the film**, which is always less than 10µm. Typical values for the mass diffusion coefficients are around 4 $10^{-9}$m$^2$/s for the mixtures considered. The characteristic time for diffusion over 10µm is 0.025s, which supports acceptably the assumption. The same approximation was used when discussing the "tears of wine" geometry, and was found to work rather well [26]. Then, the local concentration depends only on distance $r$ to the drop axis, and time $t$.

Let 1 be the more volatile compound, 2 the less volatile one. The local composition of the mixture can be defined as:

$$x(r,t) = \frac{n_1(r,t)}{n_1(r,t) + n_2(r,t)} \qquad (4)$$

$n_1(r,t)$ is the number of molecules 1, $n_2(r,t)$ the number of molecules 2, both per unit volume. The difference between the size of the molecules is neglected, and the mixture is ideal, which means that the two compounds evaporate independently.

Let $j_1$ and $j_2$ characterise the evaporation rate of the pure alkanes. If the concentration were homogeneous over the drop, one could write, without further assumption:

$$J(r,t) = \frac{j_1 x(t) + j_2 (1-x(t))}{R(t)\sqrt{1-\left(\frac{r}{R(t)}\right)^2}} \quad (5)$$

In reality, the concentration depends explicitly on $r$ and (5) holds only at very short times after drop deposition. **At longer times, the formula (5) becomes empirical** but probably still acceptable, because the increase of the evaporation rate at the edge of the drop is correctly accounted for in the denominator, as well as the relative weight of each alkane in the local evaporation in the numerator. Therefore, we shall use the approximation:

$$J(r,t) \approx \frac{j_1 x(r,t) + j_2 (1-x(r,t))}{R(t)\sqrt{1-\left(\frac{r}{R(t)}\right)^2}} \quad (6)$$

One may write:

$$\gamma(r,t) = \gamma_1 x(r,t) + \gamma_2 (1-x(r,t)) \qquad \nabla \gamma(r,t) = (\gamma_1 - \gamma_2)\nabla x(r,t) \quad (7)$$

with

$$\gamma_1 - \gamma_2 < 0 \qquad j_1 - j_2 > 0$$

The coupled equations in $h$ and $x$ can finally be written as:

$$\frac{\partial hx}{\partial t} + \nabla\left(hxU + J_{x,diff}\right) \approx -j_1 \, x \, f(r,R) \quad (8)$$

$$\frac{\partial h}{\partial t} + \nabla(hU) \approx -[j_1 x + j_2 (1-x)] f(r,R) \quad (9)$$

where radial mass diffusion has been included. Here:

$$f(r,R(t)) = \frac{1}{R(t)\sqrt{1-\left(\frac{r}{R(t)}\right)^2}} \quad (10)$$

The system is far from simple. As in the case of pure liquids, it is possible to integrate the conservation equation over the drop free surface to obtain the rate of change of the volume $V$. However, there is no reason why $x(r,t)$ should be a mere function of $\frac{r}{R(t)}$, which means that the integral of the rhs is no longer strictly proportional to the drop radius.

Note that **if radial mass diffusion were fast**, the problem would be simple, as the local concentration $x(r,t)$ would depend only on time $x(r,t) = x(t)$ and no gradient would come into play. Let us briefly discuss that case, later referred to as the "homogeneous droplet model".

From studies of pure alkanes, we know that to a first approximation an evaporating drop in the absence of gradients has the shape of a spherical cap, with approximately constant contact angle $\theta_0$ [18]. Integrating both sides of the conservation equation for the two components gives the couple of equations (11):

$$\frac{dR^3}{dt} = -\frac{8}{\theta_0} R \left[ j_1 x + j_2 (1-x) \right] \qquad \frac{d(xR^3)}{dt} = -\frac{8}{\theta_0} R \, j_1 \, x \qquad (11)$$

The variation $R(t)$ can be obtained numerically and is represented on **figure 8**. Whatever the value of $x_0$, the behaviour of the pure, less volatile alkane is recovered at long times. Noticeably, the slope of the log-log plot on the figure is significantly larger than 0.5 at the beginning of the receding motion, which merely means that the evaporation rate of the mixture slows down with time, as compared with the one of a pure alkane.

In the reality, the assumption of a complete mixing over the whole drop is irrelevant for the alkanes considered. We shall keep in mind that there are two characteristic length scales namely, the thickness and the lateral expansion, which are two orders of magnitude different. Therefore, even though complete mixing hypothesis is acceptable in the thickness of the drop, a characteristic time for diffusion to occur over the drop is of the order of the drops' life time (typically 250s for 1mm radius).

Thus, we may consider the opposite case, i.e., what happens when **radial diffusion is neglected**. Then, the equations become:

$$\frac{\partial hx}{\partial t} + \nabla (hxU) \approx - j_1 \, x \, f(r,R) \qquad (12)$$

$$\frac{\partial h}{\partial t} + \nabla (hU) \approx - \left[ j_1 x + j_2 (1-x) \right] f(r,R) \qquad (13)$$

**IV.2.a. Short time response to initial, homogeneous conditions**

It is useful to look at the change in local concentration merely due to the difference in volatility of the two alkanes, i.e., ignoring the hydrodynamic flow. This can be interpreted as a short time response of the system from initial, homogeneous conditions $x(r, t=0) = x_0$.

One may write:

$$\frac{\partial (hx)}{\partial t} \approx - j_1 \, x f(r,R) \qquad (14) \qquad \frac{\partial h}{\partial t} \approx -\left[ j_1 x + j_2 (1-x) \right] f(r,R) \qquad (15)$$

Therefore:

$$\frac{\partial x}{\partial t} \approx -\frac{f(r,R)}{h} x(1-x)(j_1 - j_2) \qquad (16)$$

The important result here is that the change in local concentration is directly controlled by the function

$$g(r,t) = \frac{f(r,R(t))}{h(r,t)} (j_1 - j_2) \qquad (17)$$

which increases with $r$ and has a steep maximum at the edge of the drop. The larger $r$, the faster the most volatile compound will be lost, and an outwardly directed surface tension gradient will take place.

$$\nabla \gamma = (\gamma_1 - \gamma_2) \nabla x \qquad (18)$$

The place where the gradient is maximum is expected to drift somewhat with time towards the centre, especially at low initial concentrations, where the most volatile alkane will completely disappear at the edge (see **figure 9**).

If one wants to write explicit formulae in the range of times where $g(r,t) \approx g(r,t=0)$:

$$x \approx \frac{x_0 \, e^{-gt}}{1 - x_0 + x_0 \, e^{-gt}} \qquad (19)$$

The characteristic response time for the local concentration is $g^{-1}$ and is quite short at the edge, where both $x$ and the gradient $\nabla x$ will rapidly vanish.
The gradient $\nabla x$ can be written as:

$$\nabla x \approx -\frac{x_0 (1-x_0) t \, \frac{\partial g}{\partial r} \, e^{-gt}}{\left(1 - x_0 + x_0 \, e^{-gt}\right)^2} \qquad (20)$$

And, if we assume an initial spherical cap with contact angle $\theta_0$:

$$g(r,t=0) = \frac{2R}{\theta_0 \left(R^2 - r^2\right)^{3/2}} (j_1 - j_2) \qquad (21)$$

**IV.2.b. Empirical discussion**

At longer times, a radial hydrodynamic flow will take place. The outwardly directed Marangoni velocity, which scales like $\left[\frac{h}{2\eta} \nabla \gamma\right]$ will depress the centre of the drop and build up a thicker rim, as observed in the experiment (figures 5-6), which means that $h(r)$ is no longer monotonous. However, flow also causes some mixing and reduces the concentration gradients.

What happens next is hardly predictable but can be explained a posteriori.

- If $x_0$ is small, which is the case in the $\alpha$-domain of the diagram, the gradient fades out with time because the most volatile alkane disappears. The drops recover progressively a spherical cap shape. As previously explained, the boundary of the $\alpha$-domain depends both on initial concentration and drop size because capillary pressure and mass diffusion, which are restoring terms, are important for small drops.

- If $x_0$ is somewhat larger, two processes come into play. First, the most volatile alkane disappears at the edge, which shifts the maximum gradient inside the drop, contributing to thin the film at that place. Second, there is a direct coupling between local concentration and local thickness $h(r)$ because $\dfrac{\partial x}{\partial t}$ scales roughly like $\dfrac{f(r,R)}{h}$. If the restoring terms are not large enough, this could lead to a local dewetting of the substrate, as observed in the experiments. This qualitative analysis could account for the $\gamma$-domain of the diagram.

- In $\beta$-domain, the thinner zone in the drop shifts progressively to the edge, and reaches it before drop breaking. For very high concentration (~80%), the rim disappears before the drop has reached its maximum extension. For lower concentration, this is rather the edge of the drop which recedes fast enough to merge with the thin zone. The main feature in $\beta$-domain is the large value of flows, which corresponds to the short lifetime of the drops. Contrary to the case $x_0 \ll 1$, where the gradients basically fade out with time, the drops where $x_0 \approx 1$ will cross the domain of large gradients ($x_0 \approx 0.5$) during their life.

**IV.2.c. Further comments about the dynamics of stable drops**

A common feature of our drops is the fast spreading rate, as illustrated in figure 1c. The initial spreading rate is approximately $R(t) \propto t^{1/10}$, as for a pure, non-volatile, wetting liquid. After a few seconds, the gradients come into play, while the drop volume starts to decrease. The net result of gradients and evaporation is a faster spreading, $R(t) \propto t^m$. The exponent m is found between 0.2 and 0.3 depending on drop size and composition and it is practically constant till the drop reaches its maximum extension.

The surface tension gradient is directed outwards during spreading, which increases the radius spreading rate. Although the rate of loss of volume increases accordingly, the maximum radius is always larger in the presence of gradients, at least for the mixtures considered.

For low concentrations ($\alpha$ domain), the slope $y$ of the log-log plot is larger than 0.5 in the first part of the retraction. As seen previously in the homogeneous drop model, such a slope is not necessarily the signature of an inwardly directed surface tension gradient, but it provides evidence of the presence of a small amount of a more volatile compound in the liquid.

At larger initial concentrations, figure 3 shows that the drops are still mixtures at the end of their life. Like in the previous case, the gradient is certainly directed outwards till the beginning of the retraction. As the slope $y$ is less from 0.5 in the first part of the retraction, one might assume that it is still the case. Why $y$ becomes close to 0.5 during the last part is possibly fortuitous. The question is open.

**Conclusion**

The present paper presents the various scenarios possible for evaporation of drops of binary alkane mixtures, depending on drop volume and mixture concentration. As the two components have different surface tensions, surface tension gradients show up rapidly and control the dynamics of the system. Noticeably, the problem is not symmetrical with respect to the two components, and only the case where the most volatile compound is in small quantity is qualitatively understood. This situation has practical interest, because it corresponds to a contamination of a liquid by a more volatile compound, with a lower surface tension. This can be detected when the drop is close to its maximum extension: the dynamics of spreading is accelerated, and so is the dynamics of retraction if compared with the one at larger times.

**Acknowledgements**

We kindly recognise the part taken by Pierre Artola in the experimental study.
**References**

[1] P.C.Wayner, Langmuir **9**, 924 (1993)

[2] I.Y.Kim and P.C.Wayner, J.Thermophys. Heat Transfer **10**, 320 (1996)

[3] S.Miladinova, S.Slavtchev, G.Lebon, J.C.Legros, J.Fluid Mech. **453**, p.153 (2002)

[4] O.A.Kabov, J.C.Legros, I.V.Marchuk, B.Sheid, Fluid Dynamics, **36**, 521 (2001)

[5] O.A.Kabov, B.Sheid, I.A.Sharina, J.C.Legros, Int.J.Therm.Sci **41**, 664 (2002)

[6] D.H.Bangham, Z.Saweris, Trans.Faraday Soc. **34**, p.554 (1938)

[7] W.D.Bascom, R.L.Cottington, C.R.Singleterry, Adv.Chem. Ser. **43**, p.355 (1964)

[8] V.X.Nguyen, K.J.Stebe, Phys.Rev.E **88**, p.16 (2002)

[9] M.H.Adao, A.C.Fernandes, B.Saramago, A.M.Cazabat, Colloids and Surfaces, **132**, p.181, 1998

[10] Y.O.Popov, T.A.Witten, Eur.Phys.J.E, **6**, 211 (2001)

[11] C.Poulard, G.Guéna, A.M.Cazabat, A.Boudaoud, M.Benamar, Langmuir **21**, p.8226 (2005)

[12] A.Marmur, M.D.Lelah, Chem.Eng.Commun., **13**, p.133 (1981)

[13] D.Pesach, A.Marmur, Langmuir **3**, p.519 (1987)

[14] R.Vuilleumier, V.Ego, L.Neltner, A.M.Cazabat, Langmuir **11**, p.4117 (1995)

[15] X.Fanton, A.M.Cazabat, Langmuir **14**, p.2554 (1998)

[16] J.Thomson, Phil.Mag. **10**, p.330 (1855)

[17] J.B.Fournier, A.M.Cazabat, Europhys.Lett.**20**, p.517 (1992)

[18] (a) M.Cachile, O.Bénichou, A.M.Cazabat, Langmuir **18**, 7985 (2002); (b) M.Cachile, O.Bénichou, C.Poulard, A.M.Cazabat, Langmuir **18**, 8070 (2002); (c) C.Poulard, O.Bénichou, A.M.Cazabat, Langmuir **19**, 8828 (2003)

[19] N.A.Fuchs "Evaporation and droplet growth in gaseous media" R.S.Bradley Ed. Pergamon Press (1959)

[20] A.Frohn and N.Roth "Dynamics of droplets" Springer Editions (2000)

[21] B.V.Derjaguin, N.V.Churaev, V.M.Muller, "Surface forces" Consultant Bureau, New York (1987)

[22] R.D.Deegan, Phys.Rev.E **61**, 475 (2000)

[23] (a) R.D.Deegan, O.Bakajin, T.F.Dupont, G.Huber, S.R.Nagel, T.A.Witten, Nature (London) **389**, 827 (1997); (b) R.D.Deegan, O.Bakajin, T.F.Dupont, G.Huber, S.R.Nagel, T.A.Witten, Phys.Rev.E **62**, 757 (2000)

[24] J.D.Jackson, *"Classical Electrodynamics"* 2$^{nd}$ Ed. Wiley. New-York (1975)

[25] J.Crank, *"The mathematics of Diffusion"* 2$^{nd}$ Ed. Oxford University Press (1975)

[26] A.de Ryck, J.Coll.Int.Sci. **209**, 10 (1999)
12

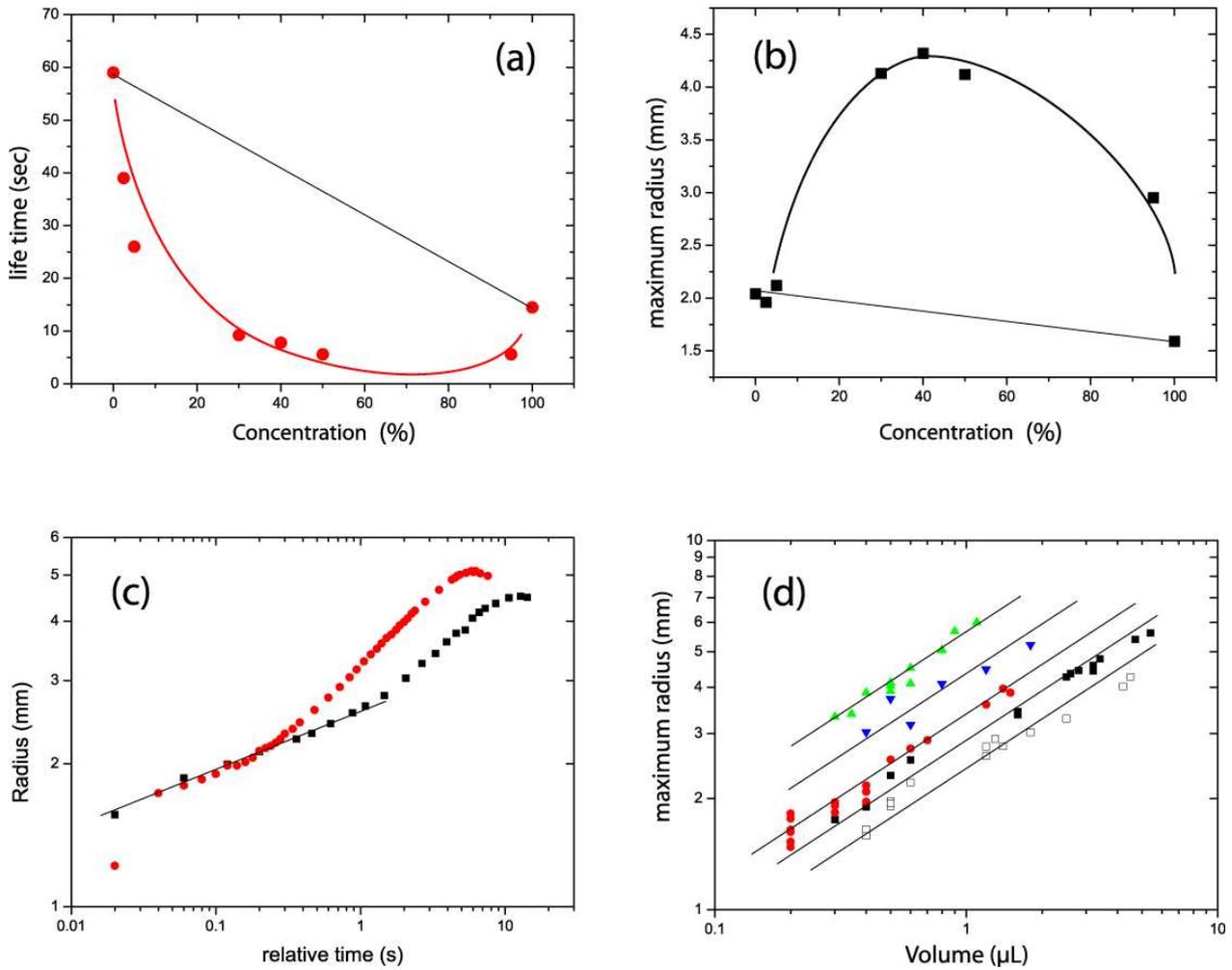

**Fig 1 :** **(a)** Life time and **(b)** maximum radius of heptane-octane mixture drops versus heptane concentration. Initial drop volume $V_o$ = 0.4µL.

**(c)** Log-Log plot of the radius versus elapsed time during the spreading stage for two drops with same initial volume Vo = 0.8µL but different concentration of heptane in octane : (■) 10%; (●) 50%. At short times, the data can be fitted by a power law (straight line) with slope 0.1. Later on, the slope increases.

**(d)** Log-Log plot of the maximum radius of the drop versus initial volume for different initial concentrations : (■) 0%, pure octane (□) 100% pure heptane (●) 2,5% (▲) 25% (▼) 95% of heptane in octane.

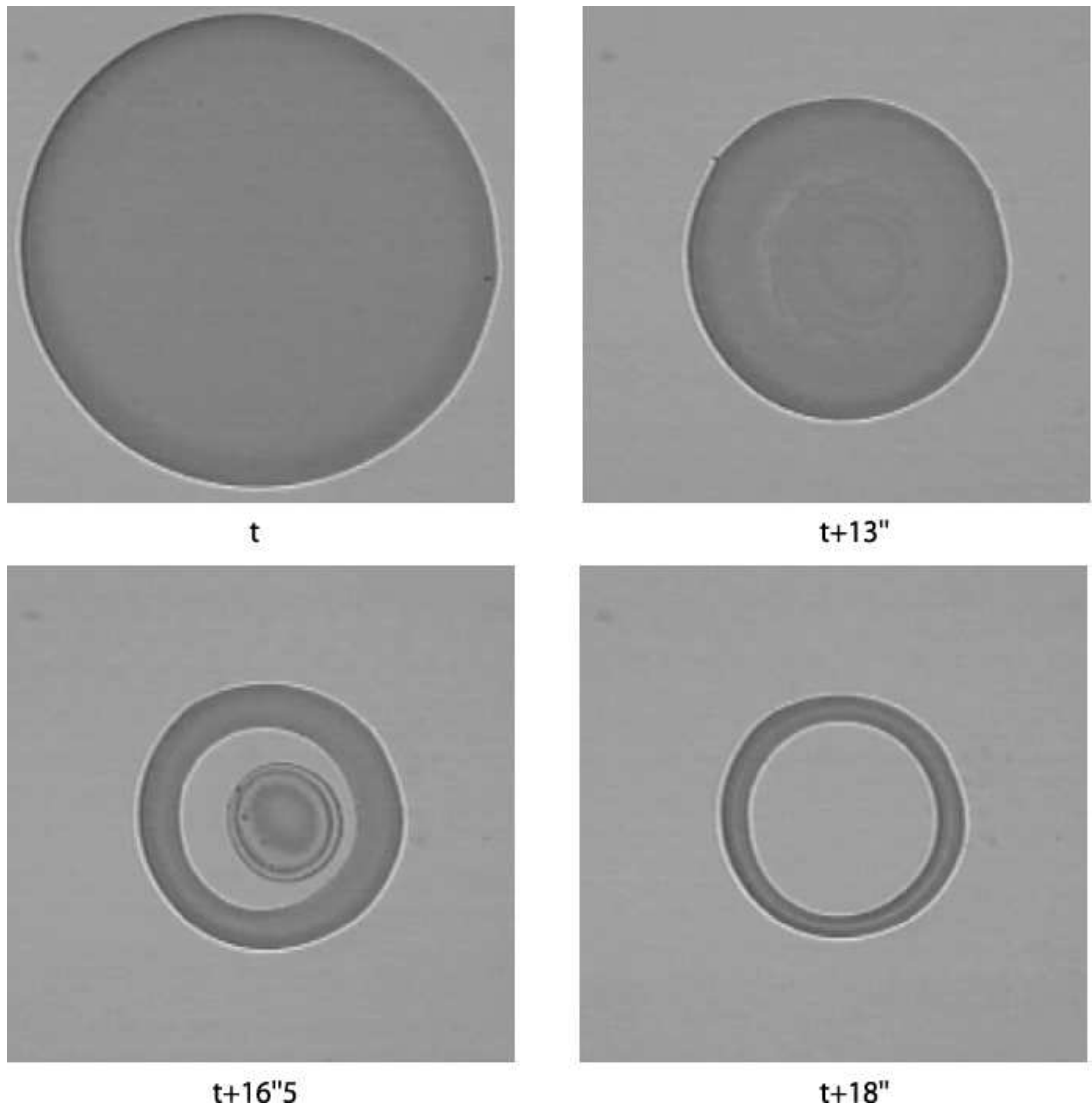

**Fig 2a :** Snapshot of an unstable drop where the central part evaporates before the outer ring. The range of unstable drops will be referred to as the $\gamma$-regime. The time t corresponds to the maximal extension of the drop. <u>Initial concentration</u> : 7% of heptane in octane. <u>Initial volume</u> : $V_o = 0.4\mu L$.

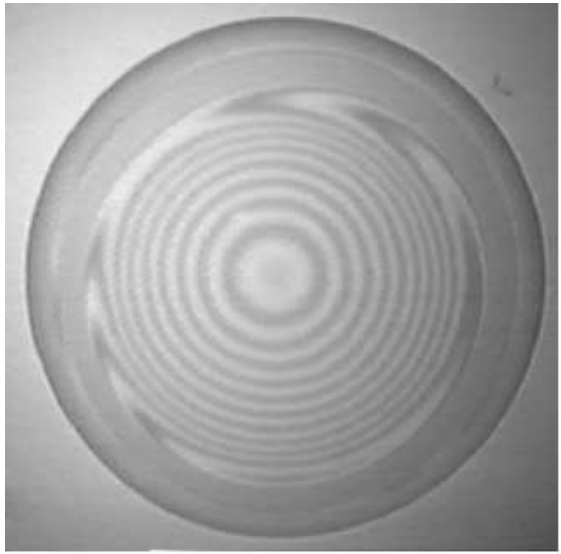 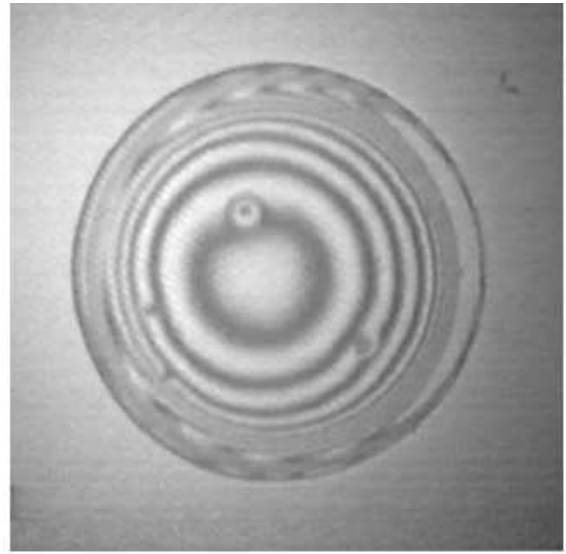

t

t + 8"

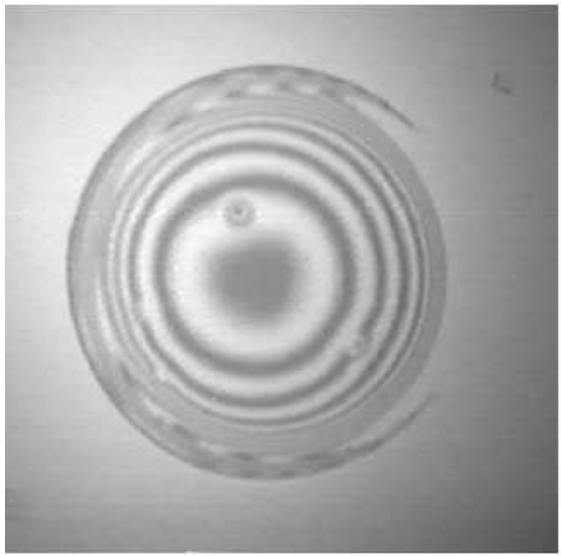 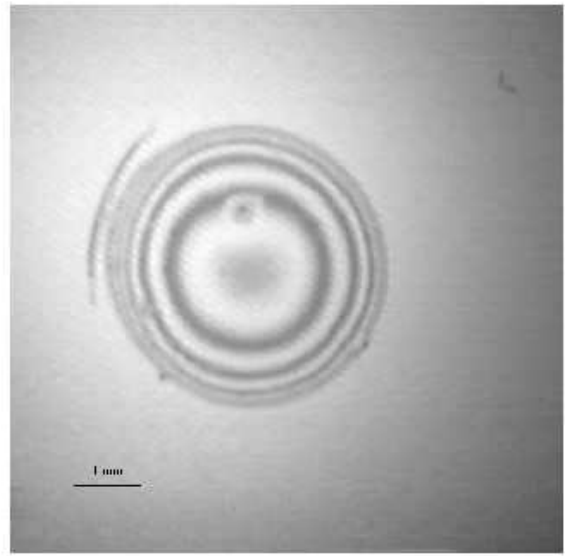

t + 8"5

t +10"

**Fig 2b :** Snapshot of an unstable drop where the central part disappears after the outer ring. The range of unstable drops will be referred to as the $\gamma$-regime. The time t corresponds to the maximal extension of the drop. <u>Initial concentration</u> : 15% of heptane in octane. <u>Initial volume</u> : $V_o$ = 0.6µL.

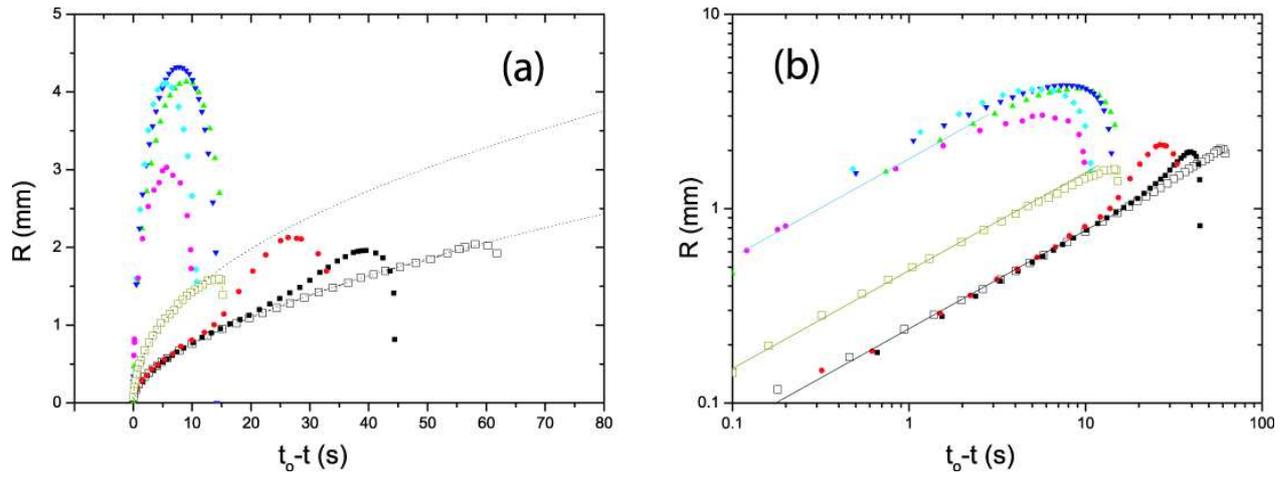

**Fig 3 :** **(a)** Linear plot of radius $R$ as a function of time before vanishing $t_0 - t$, for heptane-octane mixture drops with various concentration but same initial volume : $V_o$ = 0.4µL.

Symbols : (□) pure compound (■) 2,5% (●) 5% (▲) 30% (▼) 40% (◆) 50% (●) 95% of heptane in octane.

**(b)** Same data as in (a) in Log-Log representation.

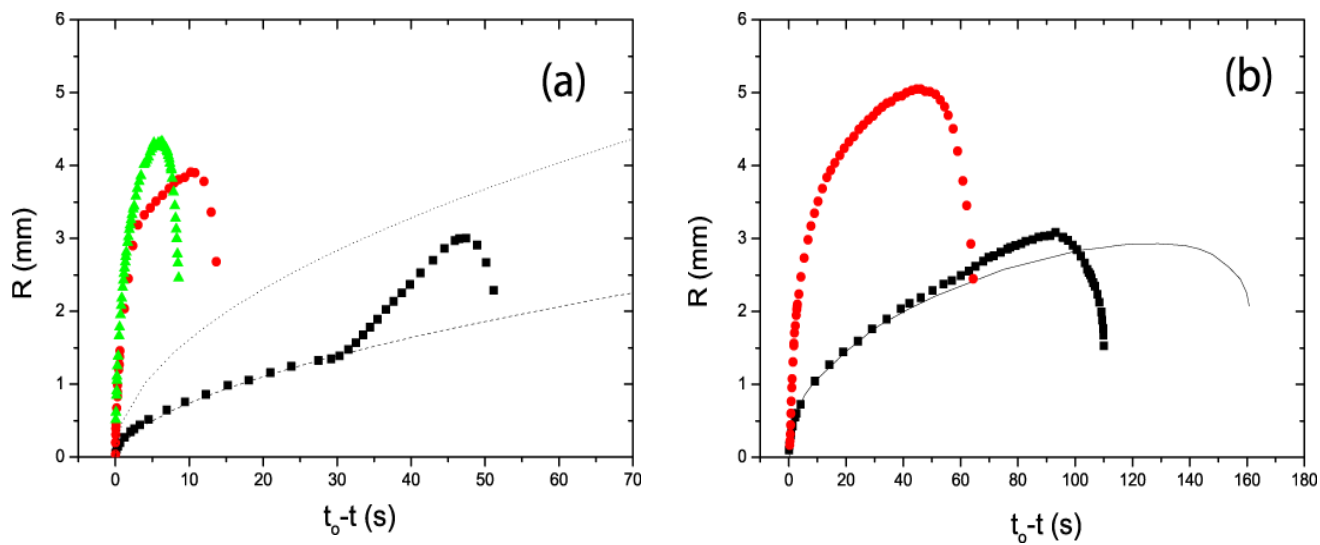

**Fig 4 :** Linear plot of radius as a function of time before vanishing for hexane-octane **(a)** and heptane-nonane **(b)** mixture drops with various concentration but same initial volume in each case.

**(a)** <u>Initial volume</u> : $V_o$ = 0.4µL (■) 5% (●) 25% (▲) 50% . <u>Dashed lines</u> : trends for pure compounds.

**(b)** <u>Initial volume</u> : $V_o$ = 0.6µL. (■) 2% (●) 10% . <u>Continuous line</u> : experimental data for nonane.

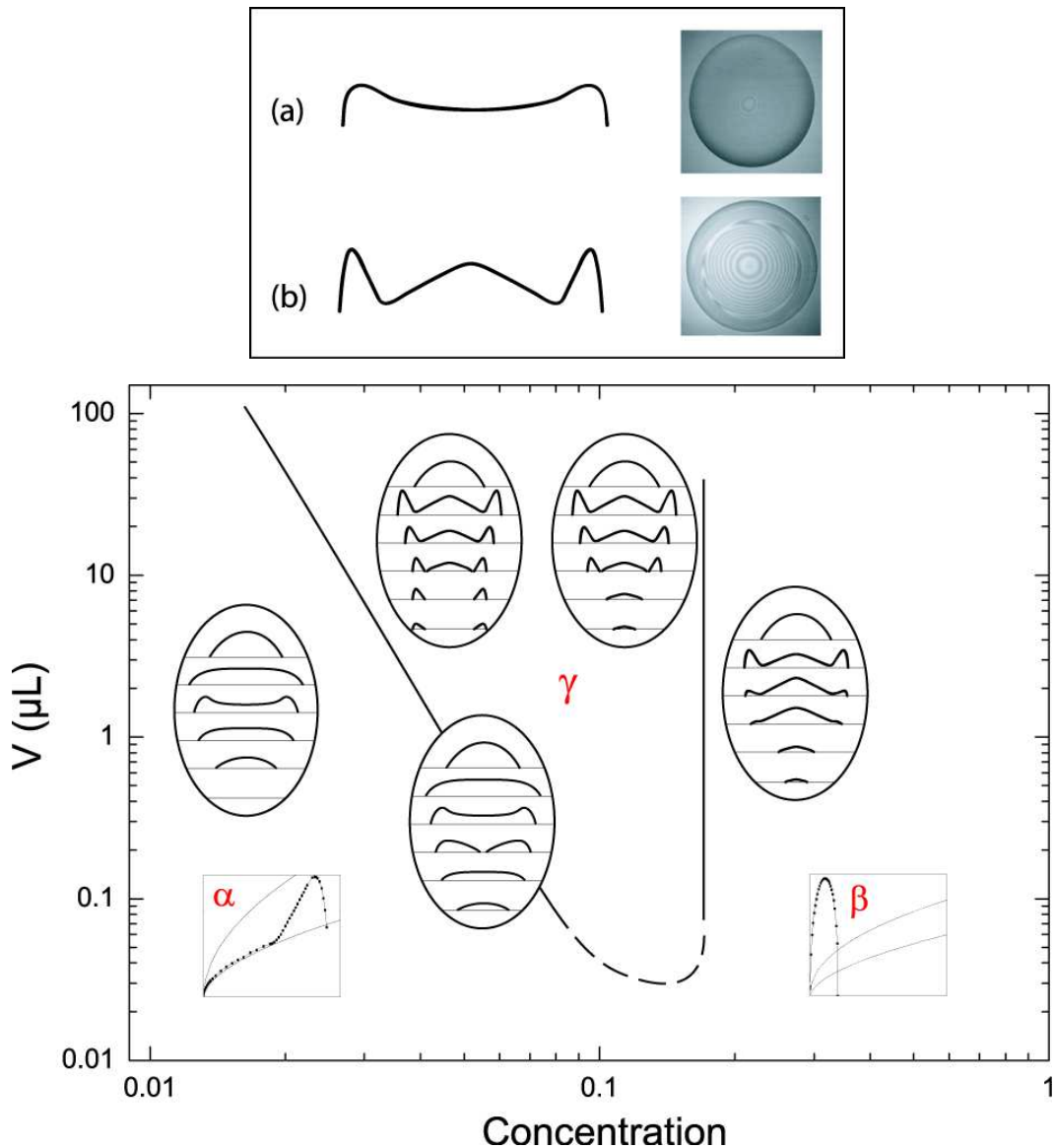

**Fig 5 :** Behaviour of alkane mixtures, as a function of concentration and drop volume

**Upper frame :** characteristic profiles reconstructed from equal thickness interference fringes. The drop may have one extremum (not shown, this is the case of a spherical cap), or two **(a)** or three **(b)** local extrema.

**Lower frame :** "Phase diagram" showing the location of $\alpha, \beta$ and $\gamma$-regimes as a function of initial volume and concentration (numerical values for axes are taken for the case of heptane-octane mixture but the diagram gives the general tendency for volatile alkanes mixtures).

Ellipses are sketches of the evolution of the drop's shape with time. They must be read from the top to the bottom. The dashed line is for the region with has not been explored systematically.

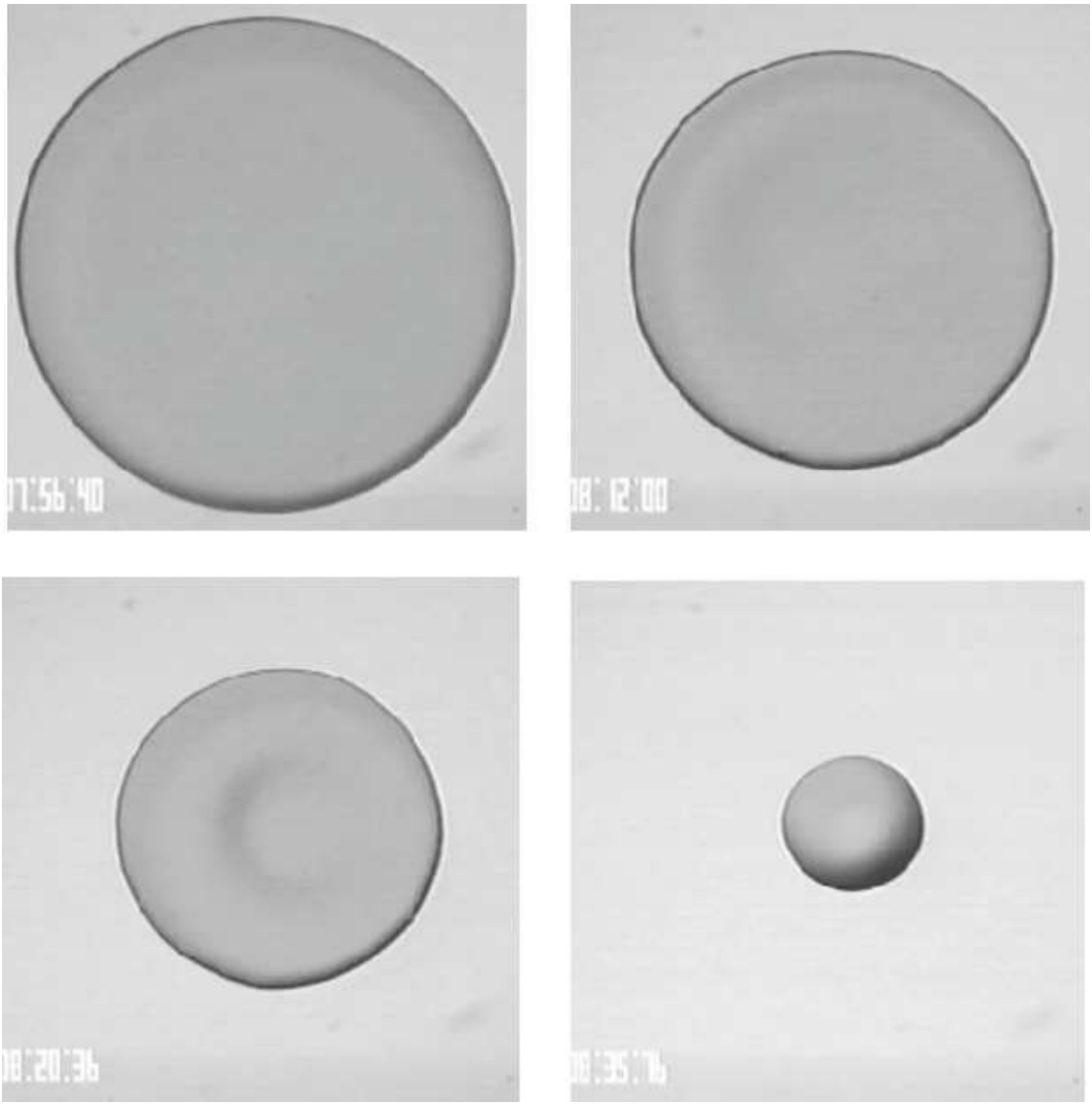

**Fig 6 :** Snapshots of a drop in the $\alpha$-regime. Initial concentration : 5% of heptane in octane. Initial volume : $V_o$ = 1.0µL. Time interval between successive pictures t ~ 15sec, the first picture is taken at the maximal extension of the drop.

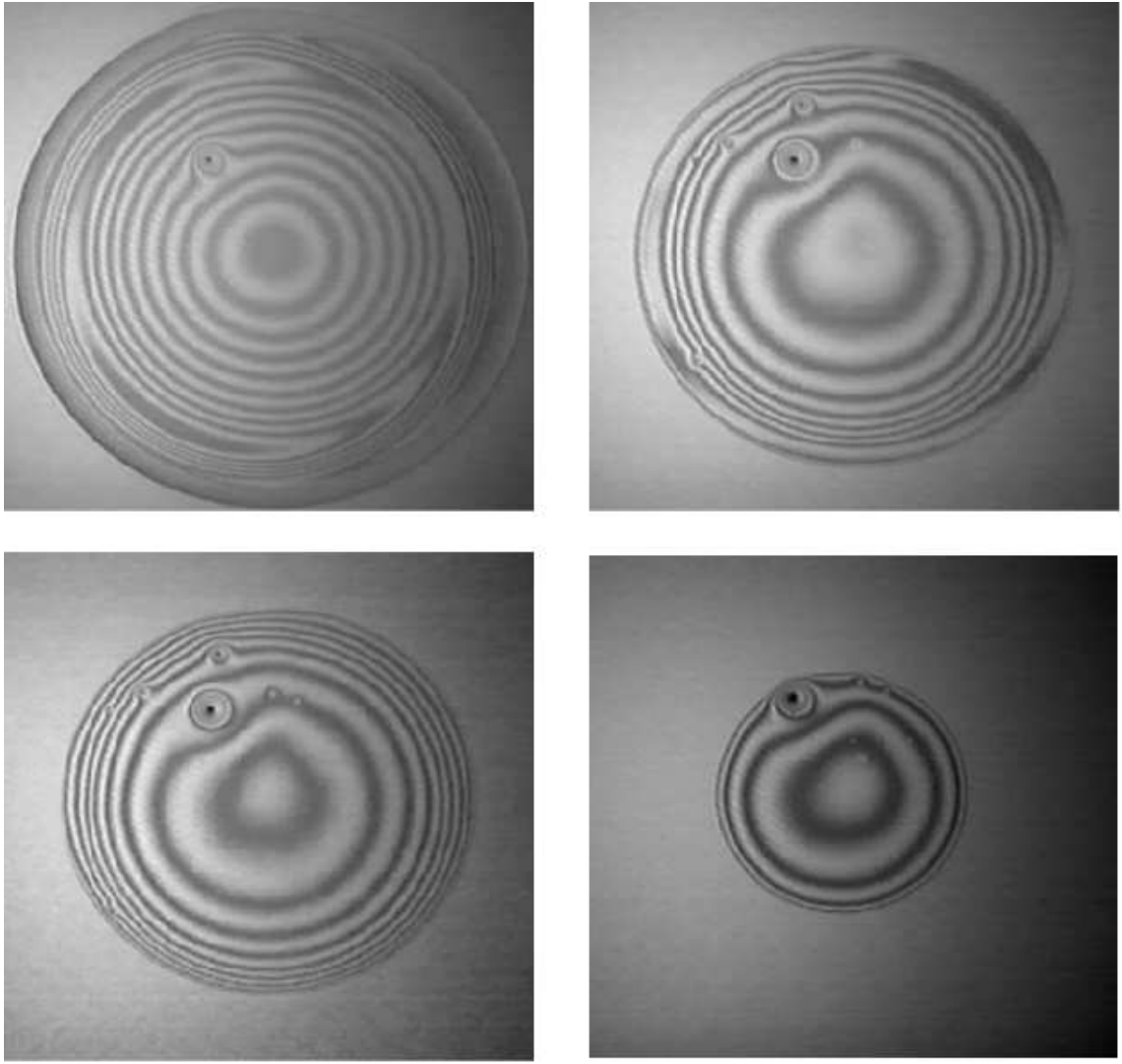

**Fig 7 :** Snapshots of a drop in the $\beta$-regime. Initial concentration : 17.5% of heptane in octane. Initial volume : $V_o$ = 0.6µL. Time interval between two pictures : t ~ 3 sec. The first picture is taken at the maximum extension of the drop.

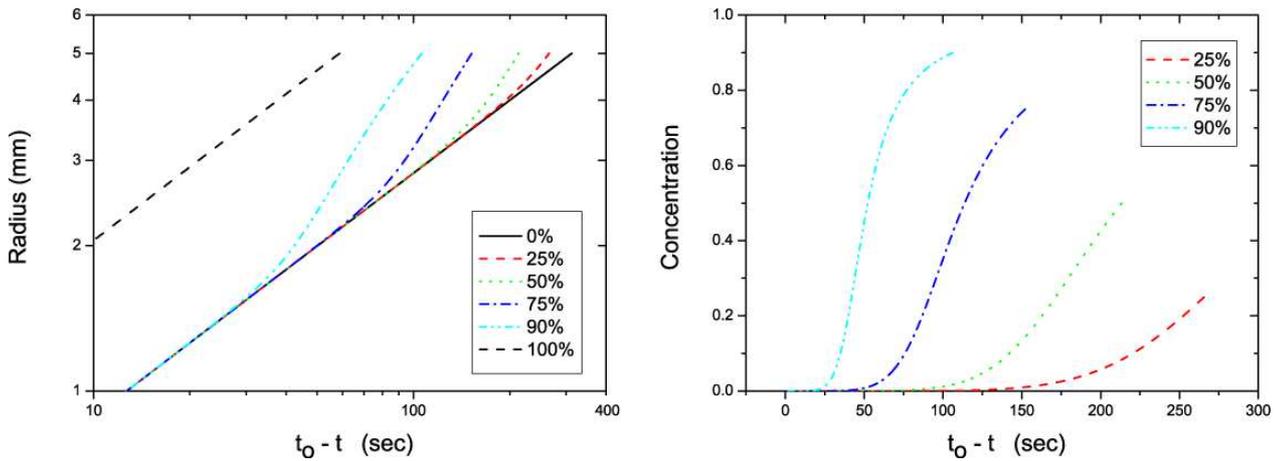

**Fig 8 :** Homogeneous droplet model: solutions of the couple of equations (11) for a homogeneous mixture of heptane in octane varying the initial concentration.

The initial radius is $R(t=0)$ = 5mm. The contact angle is supposed to be a constant $\theta$ = 0.02 rad. The evaporations parameters are respectively $j_1$ = 16.10$^{-4}$mm$^2$/s, $j_2$ = 3.10$^{-4}$mm$^2$/s. The numerical integration is performed with time increment equal to 0.01 sec until disappearance of the drop

**Left:** Log-Log representation of radius versus time interval $t_0 - t$. The curves for pure compounds are straight lines with slope 0.5. After some time, the curves for the mixtures merge with the one of the less volatile pure compound.

**Right:** Linear plot of concentration versus time interval $t_0 - t$. As expected, the life time of a given drop is a linear function of $x_o$.

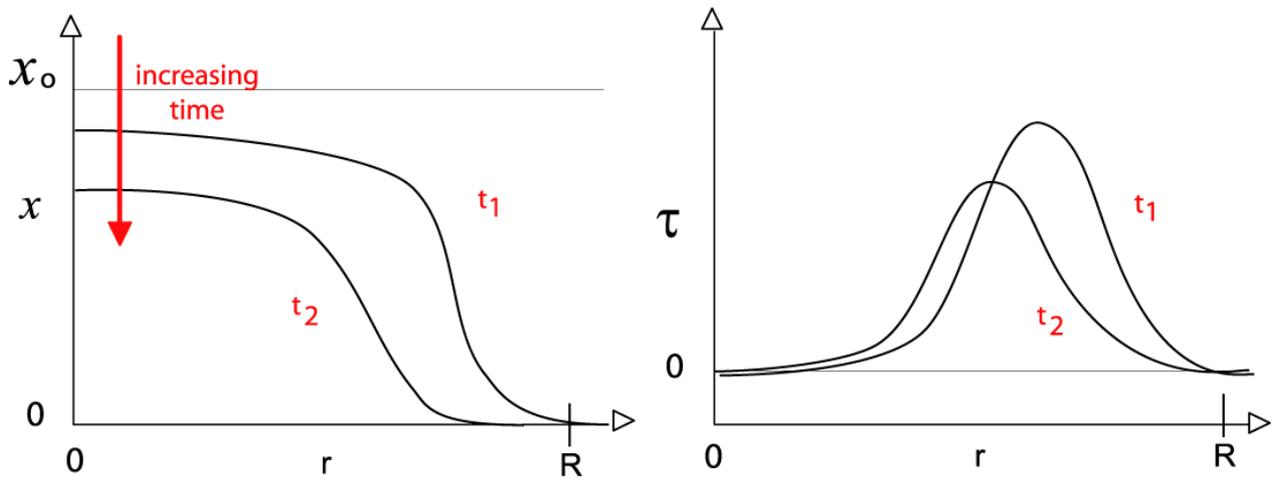

**Fig 9 :** Schematic evolution of the concentration profile and the associated shear $\tau$ at the interface. With respect to our definition, $\tau > 0$ is associated to an outwardly directed shear.